\documentclass[letterpaper, 10 pt, conference]{ieeeconf}  

\IEEEoverridecommandlockouts                              

\usepackage{amsmath,graphicx}
\usepackage{bbding}
\usepackage{amssymb}
\usepackage{multirow}
\usepackage{caption}
\usepackage{subcaption}
\usepackage{indentfirst} 
\usepackage{url} 

\title{Quadruple Augmented Pyramid Network for Multi-class COVID-19 Segmentation via CT}
\author{Ziyang Wang$^{1,*}$ 

Irina Voiculescu$^{1}$ 

\thanks{$^{*}$Ziyang Wang (e-mail: ziyang.wang@cs.ox.ac.uk) is the corresponding author.}
\thanks{$^{1}$Ziyang Wang, Irina Voiculescu are with the Department of Computer Science, University of Oxford, Oxford OX1 3QD, the United Kingdom.}

}
\begin{document}

\maketitle

\begin{abstract}
COVID-19, a new strain of coronavirus disease, has been one of the most serious and infectious disease in the world. Chest CT is essential in prognostication, diagnosing this disease, and assessing the complication. In this paper, a multi-class COVID-19 CT segmentation is proposed aiming at helping radiologists estimate the extent of effected lung volume. We utilized four augmented pyramid networks on an encoder-decoder segmentation framework. Quadruple Augmented Pyramid Network~(QAP-Net) not only enable CNN capture features from variation size of CT images, but also act as spatial interconnections and down-sampling to transfer sufficient feature information for semantic segmentation. Experimental results achieve competitive performance in segmentation with the Dice of 0.8163, which outperforms other state-of-the-art methods, demonstrating the proposed framework can segment of consolidation as well as glass, ground area via COVID-19 chest CT efficiently and accurately.
\end{abstract}

\begin{keywords}
COVID-19, Computed Tomography, Image Segmentation, Spatial Pyramid Network
\end{keywords}

\section{Introduction}
\label{sec:intro}
A novel coronal virus named COVID-19 was firstly reported in China in December 2019. The risk of COVID-19 is stated to very high at the global level by WHO in February 2020~\cite{world2020coronavirus}. More than 190 million cases have been reported in 192 countries and regions around the world, and more than 4 million patients have died until July 2021~\cite{dong2020interactive}. The spread is still ongoing. Computed Tomography(CT) is an effective medical imaging approach to diagnosis COVID-19 pneumonia, which can determine whether the patient’s lungs have lesions and the extent of the lesions. Medical image analysis plays an important role in early detection of lesions, rapid quantification, and judging whether patient is cured. In this paper, a robust and efficient medical imaging segmentation framework is proposed aiming at helping radiologists segments different regions on chest CT. The extent of damage lung volume(consolidation, glass, lung and background) is estimated properly. \\
Deep learning based computer aided diagnosed(CAD) system on COVID-19 is currently studied~\cite{zhou2020rapid}. It allows radiologists quickly get clinical information about the class of pneumonia, location, region size and etc. In image segmentation community, an encoder-decoder framework named U-Net has been an state-of-the-art segmentation method~\cite{ronneberger2015u}, and many researchers have explored COVID-19 segmentation based on it. More specifically, ~\cite{gozes2020rapid} studied on both 2D U-Net and 3D U-Net in COVID-19 segmentation, as machine learning model can collect more feature information in 3D CT, but lead to high computational cost.  \cite{chen2020deep} studied on COVID-19 segmentation mainly based on U-Net++~\cite{zhou2018unet++}. \cite{jin2020ai} build a framework based on U-Net++ which achieve jointly segmentation and classification task. Residual learning, attention mechanism, transfer learning have also been studied which improve segmentation model be more robust under noisy label, extract feature efficiently, and improving capability of the model to distinguish a variety of symptoms of the COVID-19 \cite{he2016deep}\cite{oktay2018attention}.\\
In this work, we propose a novel Quadruple Augmented Pyramid Network(QAP-Net) for multi-class COVID-19 segmentation. To solve the shortcomings of limited image features can be collected and transferred properly through multi-CNN layers pipeline, QAP-Net is proposed. Firstly, four pyramid networks are developed based on Pooling layers with variation size and Atrous CNN with variation dilation rate, respectively. Secondly, the pyramid network is theoretically improved as the augmented pyramid network. Different kinds of pooling layers and different setting of dilation rate are explored. Thirdly, four augmented pyramid networks are well analyzed and settled on a classical encoder-decoder network. Finally, QAP-Net is tested on a COVID-19 dataset which shows competitive performance against other state-of-the-art methods with a variety evaluation metrics.

\begin{figure*}[ht] 
\centering  
\includegraphics[width=\linewidth]{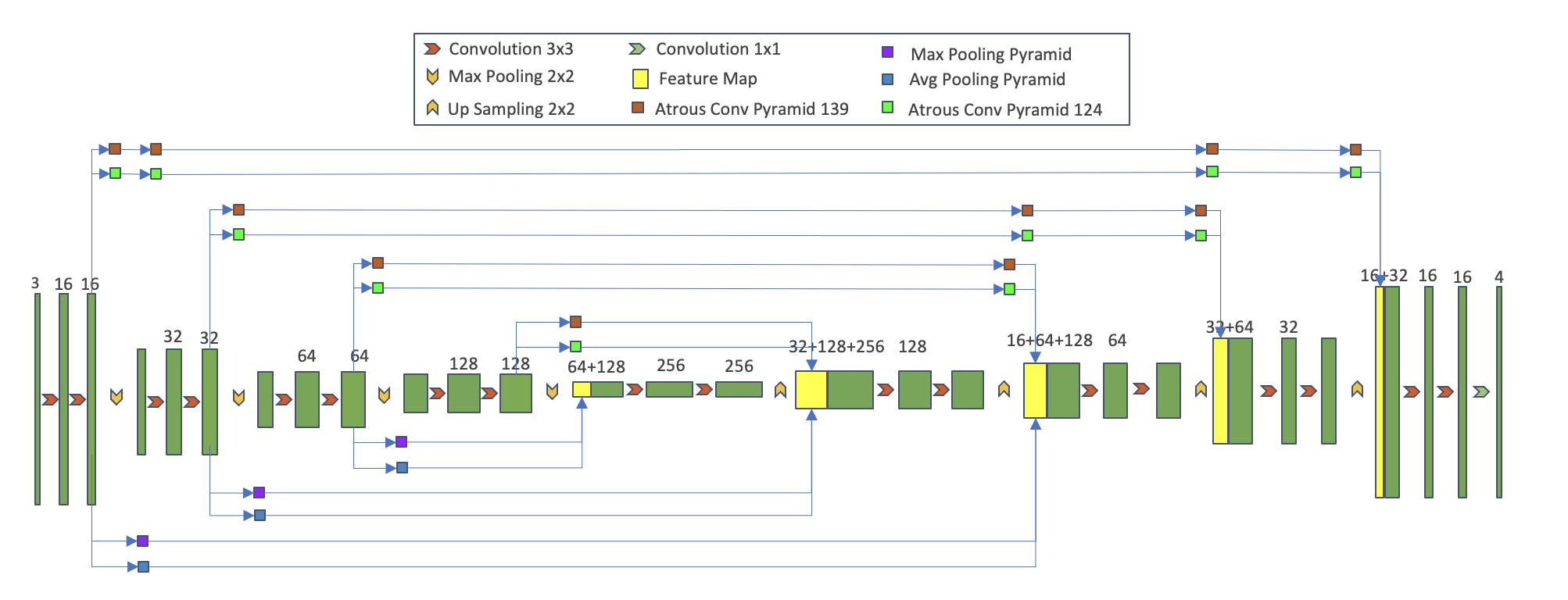}  
\caption{Quadruple Augmented Pyramid Network for Medical Image Segmentation}  
\label{fig:qapnet}  
\end{figure*}

\section{Method}
\label{sec:Method}            
The architecture of QAP-Net is illustrated in Figure \ref{fig:qapnet}. It is developed based on a classical encoder-decoder backbone including 2D Convolutioanal layers, Max pooling layers, Average pooling layers, and 2D transpose Convolutional layers. The number of channels of input, feature map and output are illustrated on each feature map. Four proposed augmented pyramid network modules showed on Figure \ref{fig:QAPmodules} are utilized on the backbone. Atrous CNN based augmented pyramid network consists of two atrous modules built in parallel approach. It performs as a skip connection from encoder to decoder, which enable sufficient global and local information been copy and crop to decoder for multi-class pixel-level segmentation. The number of these modules are 4, 3, 2, 1. Pooling based augmented pyramid network consists of Average and Max pooling modules, separately. It can extract trunk parameters, decrease computational cost, and make model more robust with rotation, translation, and size invariance. The detail of QAP-Net is introduced, analyzed and discussed below in Section \ref{sec:ACNN} and Section \ref{sec:pooling}. 
\subsection{Atrous Spatial Convolutional Network}
\label{sec:ACNN}

\begin{figure*}

     \centering
     \begin{subfigure}[b]{0.15\textwidth}
         \centering
         \includegraphics[width=\textwidth]{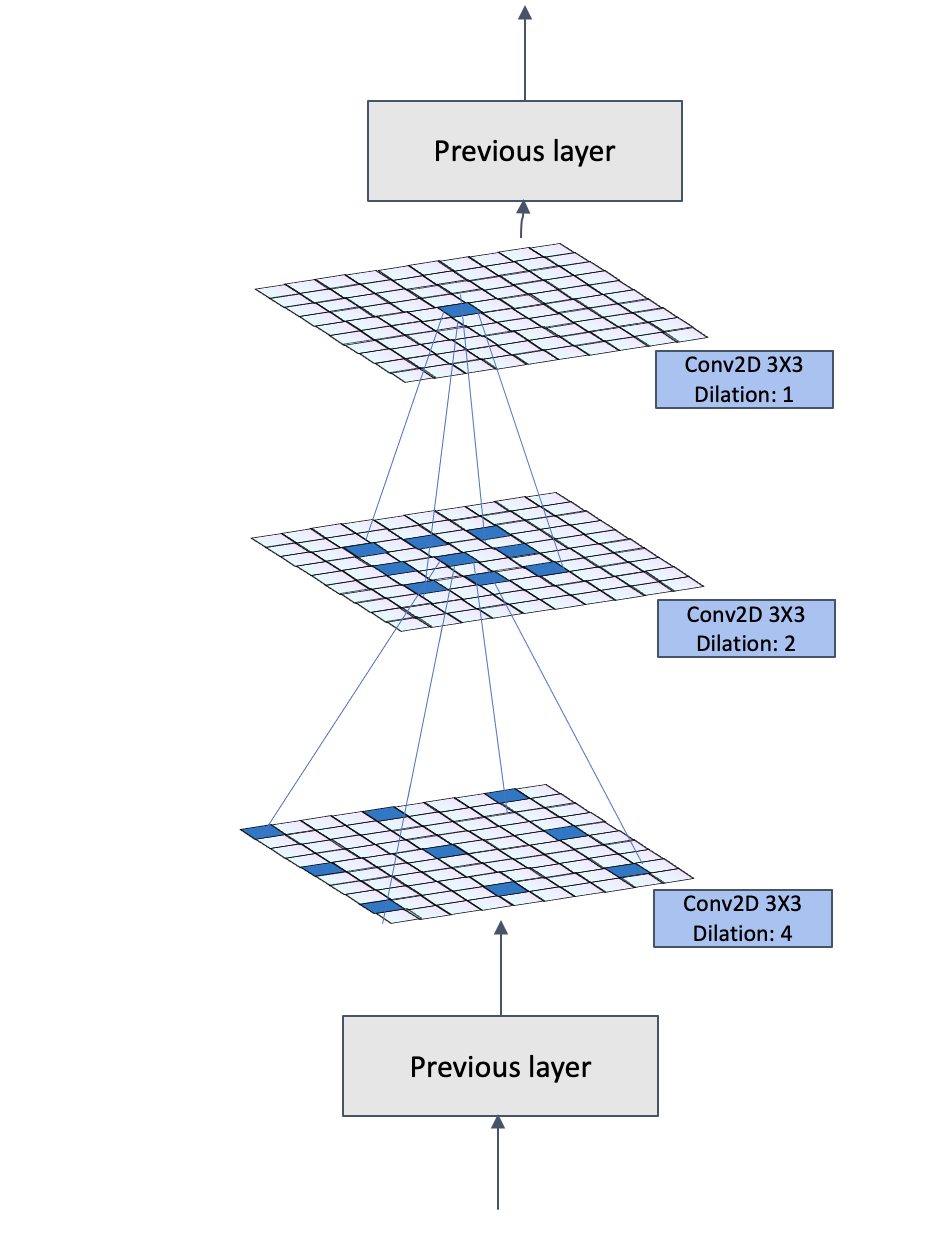}
         \caption{Pyramid Atrous Network dr=[1, 2, 4]}
         \label{fig:ACNN124}
     \end{subfigure}
     \hfill
     \begin{subfigure}[b]{0.15\textwidth}
         \centering
         \includegraphics[width=\textwidth]{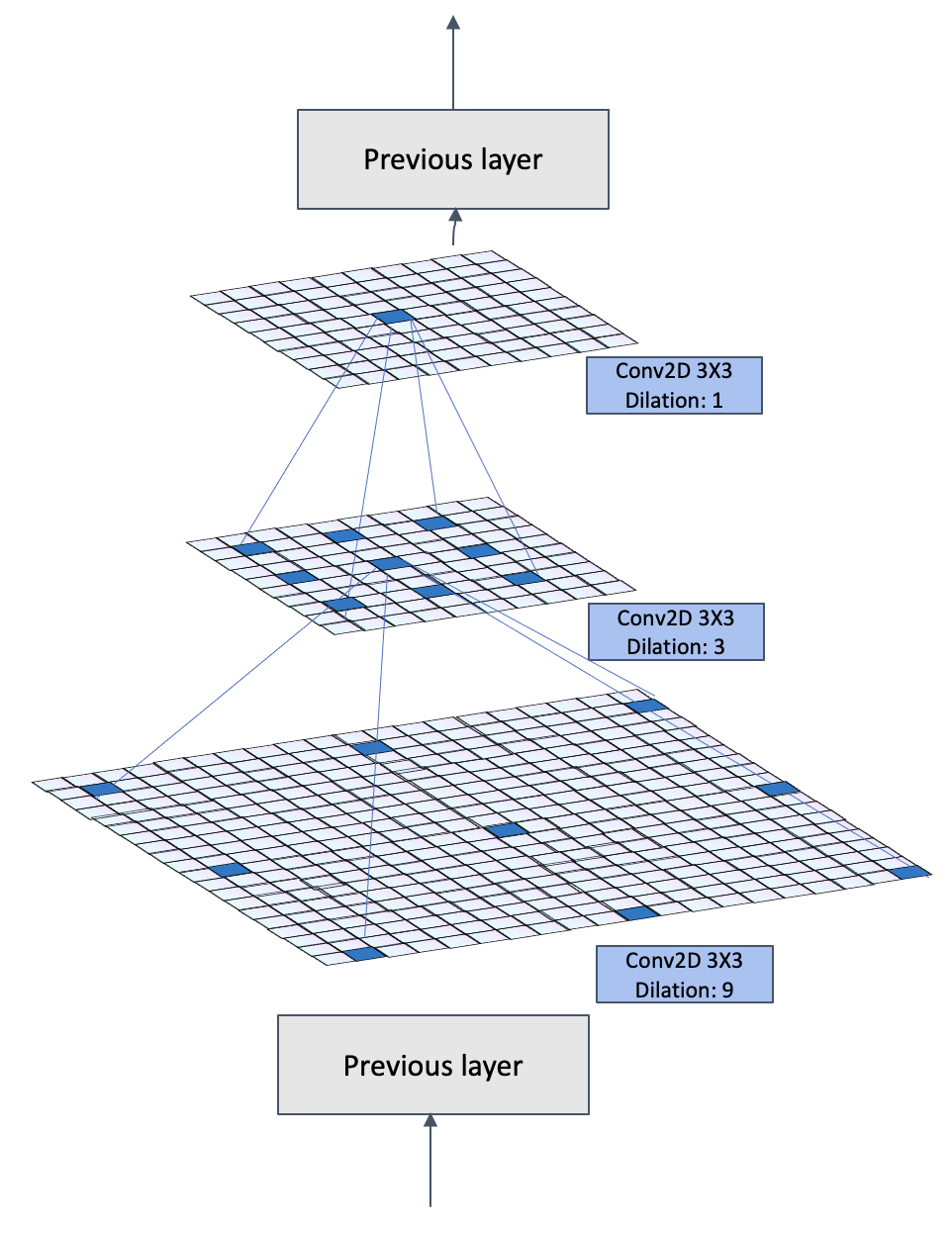}
         \caption{Pyramid Atrous Network dr=[1, 3, 9]}
         \label{fig:ACNN139}
     \end{subfigure}
     \hfill
     \begin{subfigure}[b]{0.15\textwidth}
         \centering
         \includegraphics[width=\textwidth]{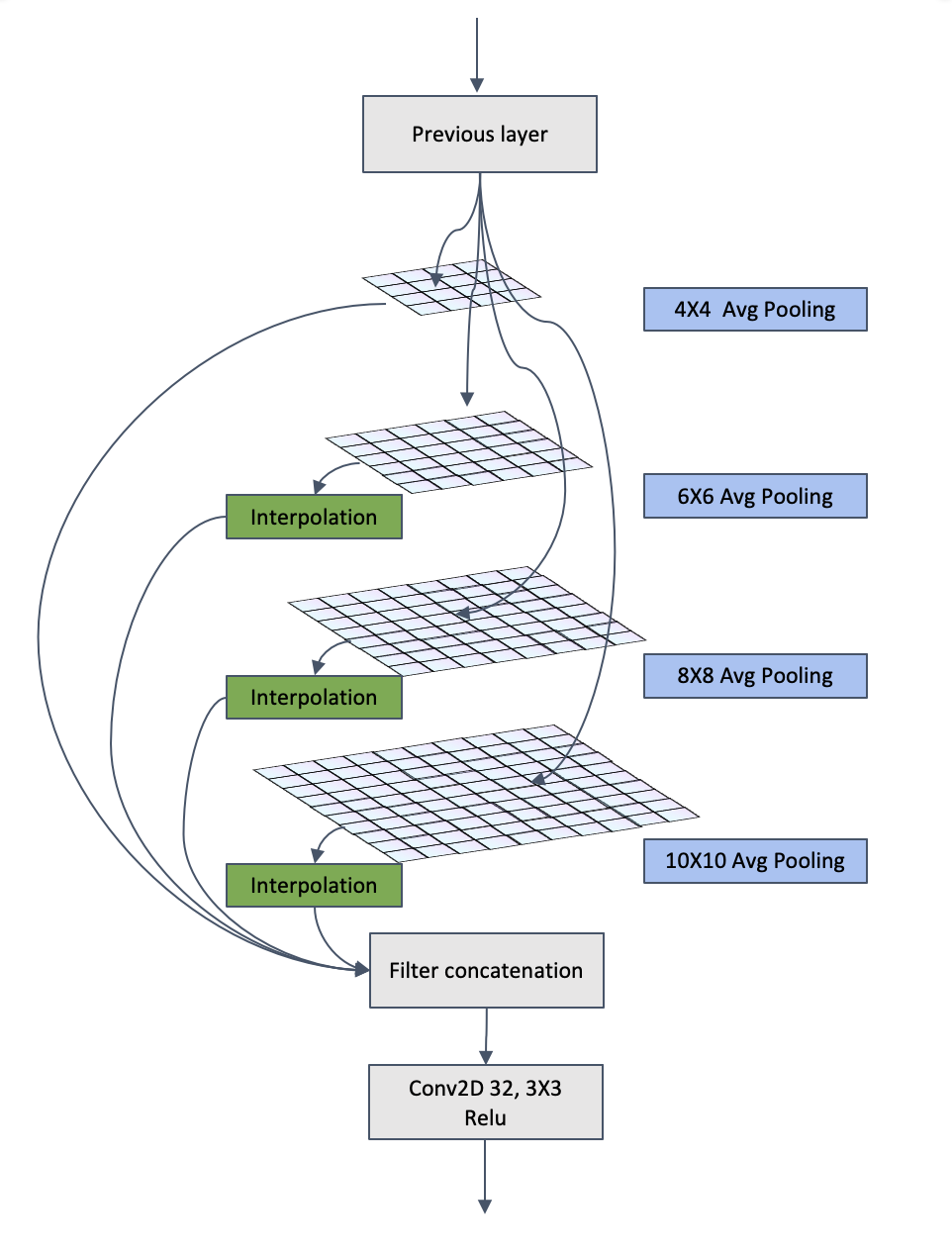}
         \caption{Pyramid Avg Pooling Network}
     \end{subfigure}
     \hfill
     \begin{subfigure}[b]{0.15\textwidth}
         \centering
         \includegraphics[width=\textwidth]{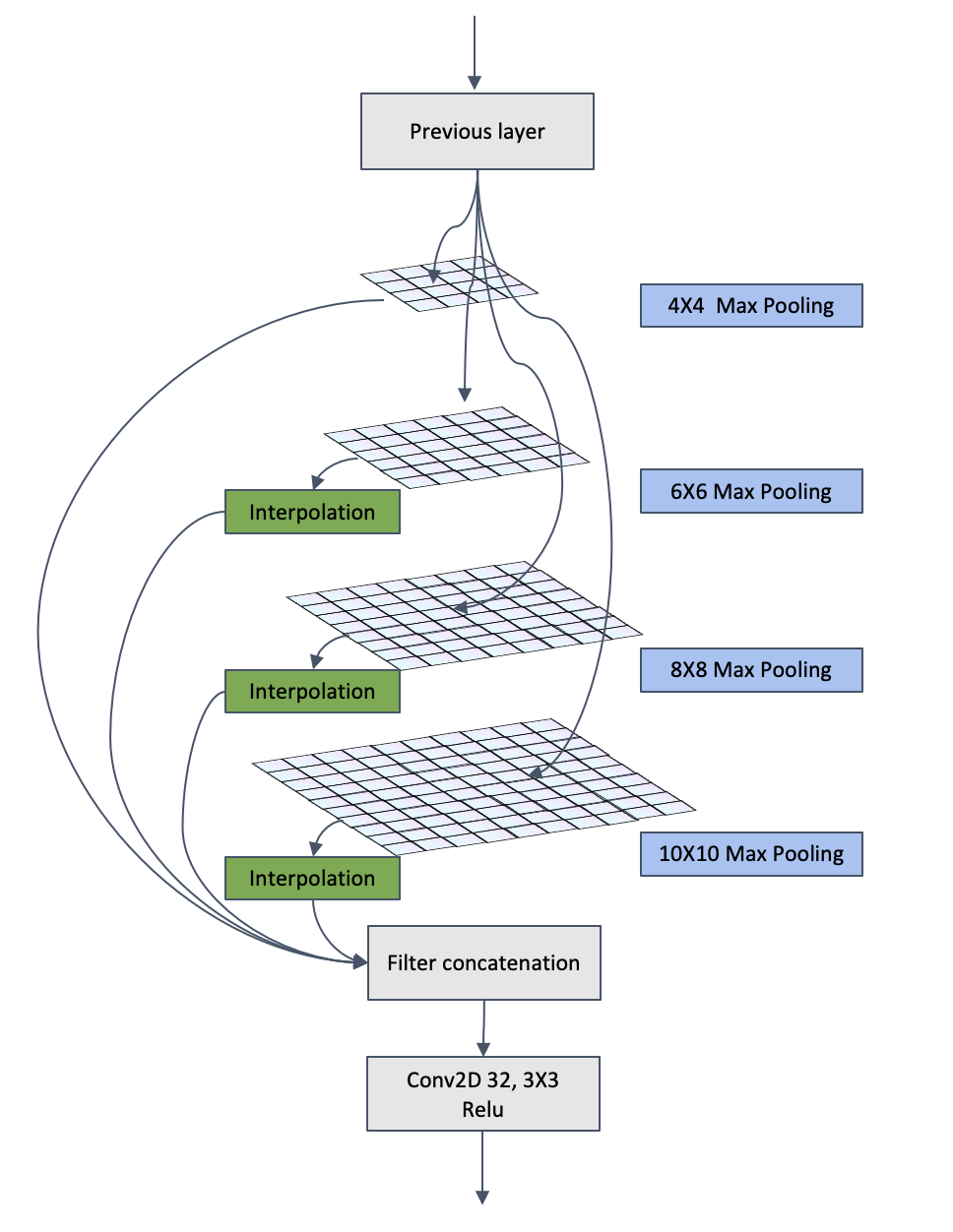}
         \caption{Pyramid Max Pooling Network}
     \end{subfigure}
     \hfill
     \caption{Augmented Pyramid Networks}
\label{fig:QAPmodules}


\end{figure*}
\begin{figure}[ht] 
\centering  
\includegraphics[width=0.9\linewidth]{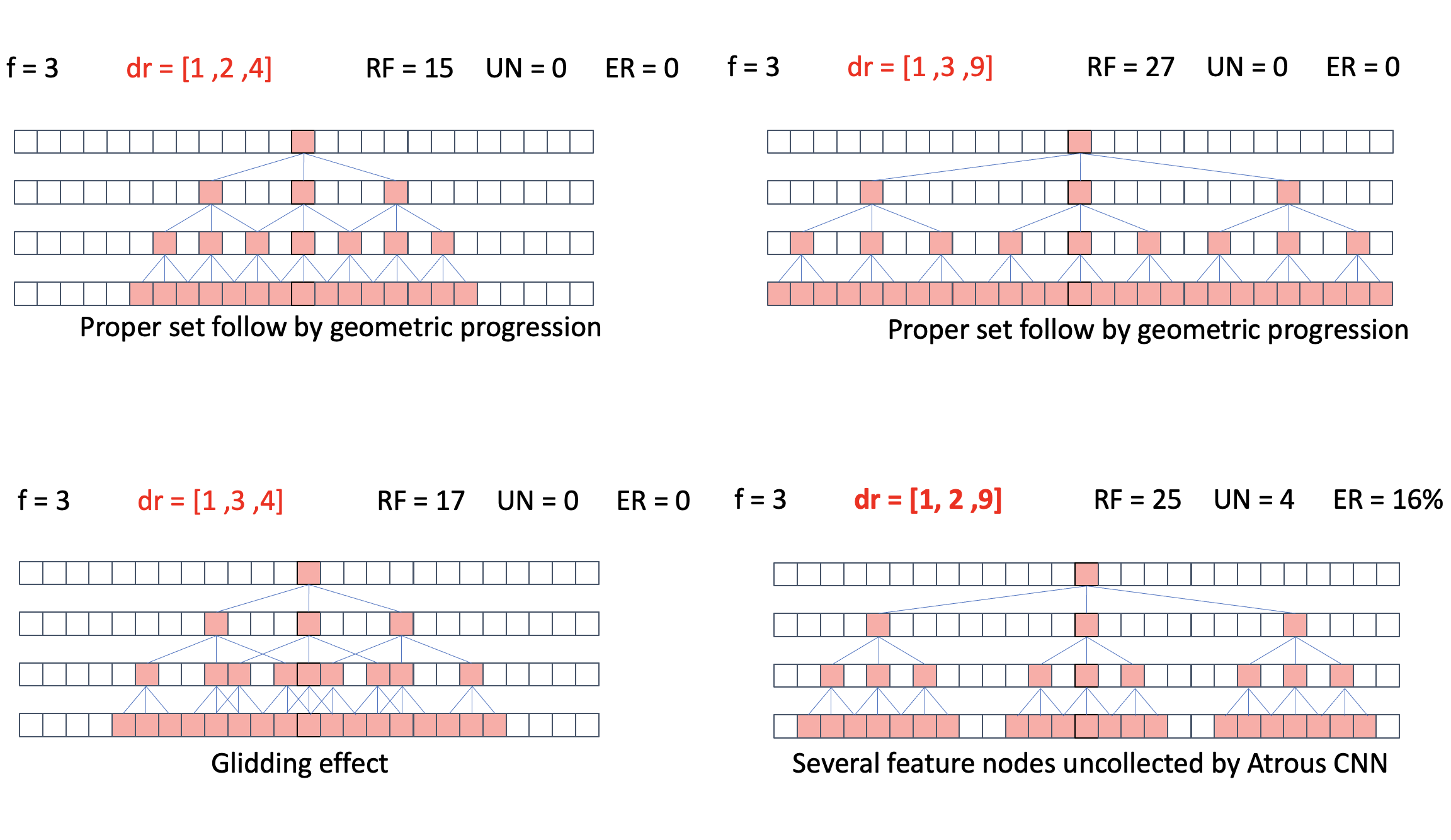}  
\caption{Example 1D Atrous CNN with Different dilation Rate Setting}  
\label{fig:ratesetting}  
\end{figure}

Convolutional Neural Networks(CNN) has shown promising performance on computer vision task in recent studies. Compared with fully connected neural networks which extract all input nodes, CNN only extract a series of  specific size of nodes on an input image also know as receptive field. Deep CNN, however, suffer not only vanishing/exploding gradients problem, but also can be detrimental to semantic segmentation task, because pixel-level features cannot been transfer through pipeline with multi-CNN layers and down-sampling. Aiming at collecting variation size of features, an alternative approach is Atrous CNN \cite{chen2017deeplab}. Atrous CNN increases receptive field by inserting zeros between non-zeros of filters. An example of atrous CNN is illustrated in Figure \ref{fig:ACNN124} and Figure \ref{fig:ACNN139}, where the dilation rate is set by 1, 2, 4 and 1, 3, 9. \\
By setting Dilation Rate $dr$ for an Filter whose size is $f * f$, the size of receptive field can be easily increased without additional computational cost. The size of Receptive Field $RF * RF$ can be calculated by Equation \ref{rfsize}.
\begin{equation}\label{rfsize}
RF = f + (dr - 1)(f - 1)
\end{equation}
The dilation rate setting, however, has not been clearly studied, and Atrous CNN lead to gridding effect \cite{yu2017dilated}. \\
To fully capture receptive field with limited number of atrous CNN by optimizing atrous rate setting, this section illustrates the reason why we develop pyramid pooling network by setting dilation rate $dr=(1,2,4)$ and $(1,3,9)$. Here, we assume the size of feature map is 1D, and no non-linear modules(Relu, Sigmpoid and etc) is utilized. $F^n$ denotes feature map is calculated by the $n^{th}$ atrous CNN, so that $F^{0}$ indicates input feature map, and $F^{n}$ indicates output feature map. The Receptive Field $RF$ of final layer can be calculated by Equation \ref{finalrfsize}.
\begin{equation}\label{finalrfsize}
RF^{n} = f + (dr^{0} - 1)(f - 1) + \sum_{1}^{n} dr^{n} 
\end{equation}
However, atrous CNN potentially lead to feature nodes not been collected as there is non-trainable zeros been inserted into filter. The number of Uncollected Nodes $UN^{n}$ after $n^{th}$ atrous CNN can be calculated by Equation \ref{UNEQUATION}.
\begin{equation}\label{UNEQUATION}
UN^{n} = (f-1)*(dr^{0} - n*\sum_{1}^{n} dr^{n}) 
\end{equation}
If $UN^{n}$ is a negative number, it means no uncollected nodes and several nodes been collacted for multi times.
To evaluate the efficiency of dilation rate setting, an Evaluation Ratio $R$ between receptive field size $RF$ and Uncollected Nodes $UN^{n}$ is calculated by equation \ref{ratio}
\begin{equation}\label{ratio}
ER =  UN^{n} / RF^{n}
\end{equation}
Under the limit and unchanged number of atrous CNN layers $n$, $ER$ should be as minimize as possible. After Equation \ref{finalrfsize} and Equation \ref{UNEQUATION} been put in Equation \ref{ratio}, the dilation rate setting $dr$ must follow geometric progression as shown in Equation \ref{setting} so that maximum the value of receptive field size $RF$, and minimum the value of Uncollected Nodes $UN^{n}$.
\begin{equation}\label{setting}
dr = [1, ... , (dr)^{n-1}]
\end{equation}
A more intuitive and visible example is shown on Figure \ref{fig:ratesetting}. Dilation rate setting such as [1,2,4] and [1,3,9] performs better than [1,3,4] which potentially leads to glidding effects or [1,2,9] which results in several feature nodes not been collected.

Inspired by Inception module~\cite{szegedy2015going}, we build an atrous CNN module based on two proper Atrous CNN pyramid networks consist of three layers. To mitigate the difference between encoder and decoder \cite{ibtehaz2020multiresunet}, and transfer more feature information to decoder, a skip-connection-liked-path is built by atrous CNN block. Considering down-sampling, up-sampling and difference between encoder and decoder, the number of atrous CNN blocks on each path is 4, 3, 2, 1.

\subsection{Spatial Pyramid Pooling Network}
\label{sec:pooling}
Pooling layers are commonly attached after convolutional layers in computer vision tasks. The main function of Pooling layers are 1)translation, rotation, scale invariance, 2) reducing computational cost by down-sampling, 3) avoiding overfitting, and 4)improving generalization ability of model. \\
Max pooling is commonly applied in image segmentation tasks, because the feature of boundary and texture structure of the image can be efficiently extracted by capturing the maximum value of pixel. Max pooling outperforms average pooling when variance of brightness of pixel is high (under noisy label, Under multi-scene conditions, Strong contrast between light and dark). Considering the application of proposed model is specialized to COVID-19 medical imaging analysis via CT. Average pooling is utilized to reduce the deviation of the estimated mean which outperforms better than Max pooling. For this reason, the final augmented pooling module is based on two sub-module with Max pooling and Avg pooling.\\
To collect variation size of features, the size of pooling operation is set to 4*4, 6*6, 8*8, 10*10, and the output by pooling layers are all resized by interpolation except the 4*4 layer, so that all features can be concatenated on channel axis. In this way, the size of feature map is four times smaller than the input feature map size. In the QAP-Net, the augmented pooling networks act as down sampling which the input is the 1st, 2nd and 3rd level of the output of encoder, and also act as skip connection that the output is concatenated as input to the related decoder, respectively.

\section{Experiments and Results}
\label{sec:pagestyle}

\subsection{Dataset and Experimental Setup}
We used a COVID-19 dataset from MedSeg, a commercial AI medical company which consists of 20 CT scans, of up to 630 slices per scan~\cite{dataset}. All images are normalized and resized to $256 \times 256$. Images for PSP-Net are resized to $288 \times 288$.
Data augmentation was applied in the form of rotations, random size crop and horizontal flip. 10\% of data is used for testing, 10\% of data is used for valdation and the rest of data is for training.
The code was developed in Python using Tensorflow. It has been run on an Nvidia GeForce RTX 3090 GPU with 24GB memory, and Intel i9-10900K. With a training batch size of 16, the learning rate is $10^{-4}$, the learning rate will be reduced by *0.8 if no dice coefficient been improved after 8 epochs, and the minimal learning rate is $10^{-5}$. Training will be early stopped, once dice coefficient on validation dataset not been improved for 10 epochs.  Given the imbalance between the spine and background pixels, the loss function was based on the categorical focal loss. The total training epoches is 100. 

\subsection{Results and Discussion}
\begin{figure}[ht] 
\centering  
\includegraphics[width=0.8\linewidth]{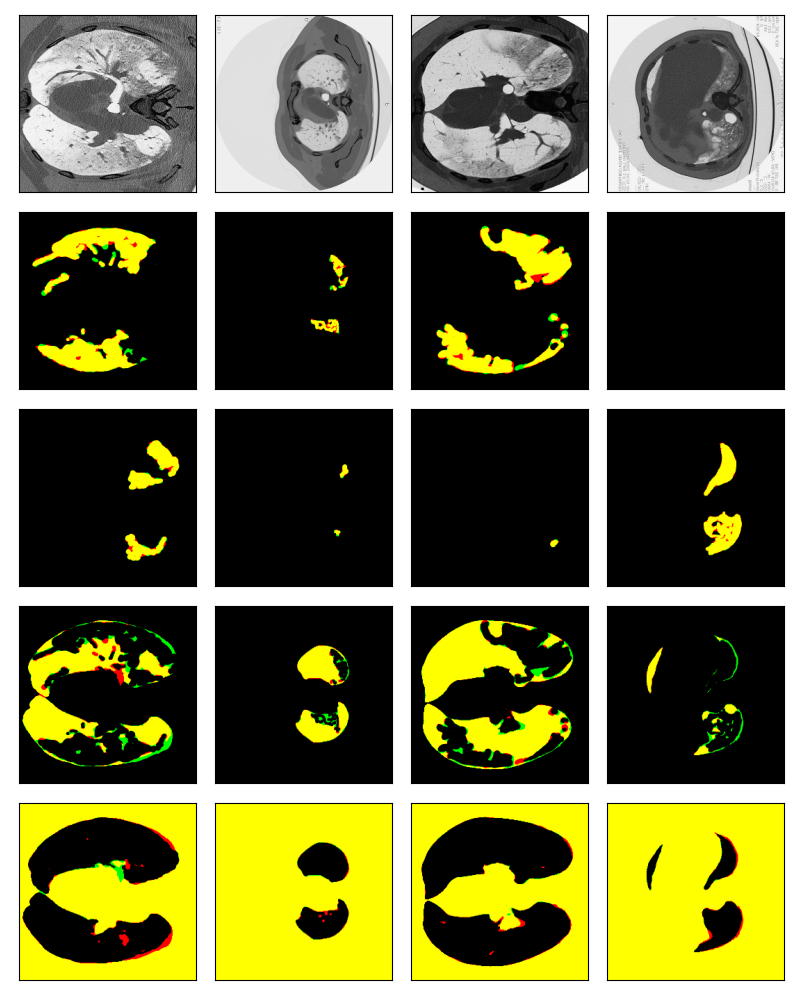}  
\caption{Example Results on Test Dataset by QAP-Net}  
\label{fig:exampleresults}  
\end{figure}

 QAP-Net is compared with classical segmentation algorithms including LinkNet, PSP-Net, MultiResUnet, DenselyUnet, and U-Net with several classical backbones such as VGG, ResNet and etc \cite{he2016deep} \cite{simonyan2014very} \cite{icip2020wang}.
 We  compare the performance of our algorithm against a collection of others conventional, widely used overlap measures such as the Dice coefficient, Accuracy, Precision, Sensitivity or Recall, Specificity, which ensuring a reliable evaluation with other methods are illustrated in Table \ref{table:overlap}. Example images about ground truth and predicted results including ground glass, consolidation, lung other and background is illustrated in Figure \ref{fig:exampleresults} where yellow demonstrate True Positive pixels, red demonstrate False Positive pixels, and green demonstrates False Negative pixels.
\begin{table}[t] \footnotesize
\begin{center}
\caption{Direct Comparison Against Existing Algorithms}
\label{table:overlap}
\begin{tabular}{llllll}
\hline\noalign{\smallskip}
Model & mIOU &  Acc & Pre & Sen & Spe  \\
\noalign{\smallskip}
\hline
\noalign{\smallskip}
LinkNet & 0.6126 & 0.9833 & 0.7181 & 0.7783 & 0.9883 \\
Resnet50 LinkNet & 0.6766 & 0.9829 & 0.7867 & 0.8110 & 0.9881 \\
VGG Linknet &0.4640 &0.8534 &0.5160 &0.6985 &0.8577 \\
Dense Linknet &0.7887 &0.9919 &0.8336 &0.9324 &0.9941 \\
U-Net & 0.7374 & 0.9894 & 0.8133 & 0.8792 & 0.9924 \\
Attention U-Net & 0.5585 & 0.8234 & 0.5871 & 0.9157 & 0.8251 \\
Efficient U-Net & 0.5828 & 0.7176 & 0.6026 & 0.9539 & 0.7180\\
Residual-U-Net & 0.7620 & 0.9895 & 0.8287 & 0.9005 & 0.9923 \\
Dense-U-Net & 0.5544 & 0.7772 & 0.5871 & 0.8921 & 0.7788 \\
MultiRes-U-Net & 0.4264 & 0.9294 & 0.4328 & 0.9757 & 0.9276 \\
RAR-U-Net & 0.7884 & 0.9969 & 0.8536 & 0.9120 & 0.9982 \\
PSP-Net & 0.7733 &0.9970 & 0.8220 &0.9260 &0.9974 \\
SeResNet PSP-Net & 0.7446 & 0.9965 & 0.7840 & 0.9365 & 0.9973 \\
QAP-Net & 0.8163 &0.9976  &0.8460  &0.9580  &0.9980  \\
\hline

\end{tabular}
\end{center}
\end{table}

\subsection{Ablation study}
In order to analyze the effects of each of the four proposed augmented networks and their combinations, extensive ablation experiments have been conducted. Table~\ref{tab:ablationcomparison} documents how the removal of one or more components compromises the overall performance. The same table also gives a measure of the complexity of the overall  model and its sub models. 
Algorithm illustrates that different algorithms are separately utilized for training.
\begin{table}[t]\small

\caption{Ablation Studies on Contributions of Architecture}
\small
\label{tab:ablationcomparison}
\begin{center}
\centering
\begin{tabular}{|p{0.16\linewidth}|p{0.16\linewidth}|p{0.16\linewidth}|p{0.16\linewidth}|p{0.1\linewidth}|}
\hline
 \multicolumn{2}{|c|}{Atrous Pyramid Net }&\multicolumn{2}{c|}{Pooling Pyramid Net} & mIOU\\
\cline{1-4}
 1 3 9 &  1 2 4 & Max  & Avg &   \\
\hline
  & &  & & 0.5331 \\
  $\checkmark$ & & & & 0.7469\\
    &    $\checkmark$  & & & 0.7966\\
   $\checkmark$    &    $\checkmark$  & & & 0.7996\\  
    &     &  $\checkmark$ & & 0.7419\\
    &     &   & $\checkmark$ & 0.7872 \\
        &     &  $\checkmark$ & $\checkmark$ & 0.7373\\
    $\checkmark$    &  $\checkmark$   &   & $\checkmark$ & 0.8072\\ 
     $\checkmark$    &  $\checkmark$   &  $\checkmark$ &  &0.8023 \\ 
        $\checkmark$    &  $\checkmark$   &  $\checkmark$ &  $\checkmark$  & 0.8163\\   
 \hline

 \hline
\end{tabular}
\end{center}
\end{table}

\section{Conclusion}
\label{sec:conclusion}
Multi-class segmentation of COVID-19 chest CT is of great significance for the diagnosis in clinical practice. We establish four augmented pyramid networks on a encoder-decoder network. Comprehensive evaluations and comparisons are completed, and our proposed method achieves promising performance.

\bibliographystyle{IEEE}
\bibliography{ref}
\end{document}